%% file: main.tex
\documentclass[10pt,conference,screen]{src/IEEEtran}
\IEEEoverridecommandlockouts
\def\BibTeX{{\rm B\kern-.05em{\sc i\kern-.025em b}\kern-.08em
    T\kern-.1667em\lower.7ex\hbox{E}\kern-.125emX}}

\input{src/packages}

\input{src/marcos}
\input{src/colors}

\newcommand\copyrighttext{%
  \footnotesize \textcopyright 2022 IEEE. Personal use of this material is permitted.
  Permission from IEEE must be obtained for all other uses, in any current or future
  media, including reprinting/republishing this material for advertising or promotional
  purposes, creating new collective works, for resale or redistribution to servers or
  lists, or reuse of any copyrighted component of this work in other works.
  Preprint of paper to be presented at IC2E 2022.}
\newcommand\copyrightnotice{%
\begin{tikzpicture}[remember picture,overlay]
\node[anchor=south,yshift=10pt] at (current page.south) {\fbox{\parbox{\dimexpr\textwidth-\fboxsep-\fboxrule\relax}{\copyrighttext}}};
\end{tikzpicture}%
}

\begin{document}
\title{\name{}: A Generalized Serverless Compute Architecture for Hardware Processing Accelerators\\\centering\large\textit{Preprint}}

\author{\IEEEauthorblockN{Sebastian Werner}
\IEEEauthorblockA{ISE, TU Berlin, Germany \\
sw@ise.tu-berlin.de}
\and
\IEEEauthorblockN{Trever Schirmer}
\IEEEauthorblockA{MCC \& ECDF, TU Berlin, Germany \\
ts@mcc.tu-berlin.de}
}

\maketitle
\copyrightnotice
\begin{abstract}
  \input{sections/00_abstract}

\end{abstract}


\section{Introduction}
\input{sections/01_introduction}

\section{Related Work}\label{sec:rw}
\input{sections/02_related_work}

\section{Problem Statement}\label{sec:prob}
\input{sections/03_problem}

\section{Design \& Implementation}\label{sec:dgn}
\input{sections/04_design_grammerly}

\subsection{Implementation}\label{sec:impl}
\input{sections/05_implementation_grm}

\section{Evaluation}\label{sec:evl}
\input{sections/06_evaluation}

\section{Conclusion}\label{sec:clu}
The increasing use of hardware processing accelerators tailored for specific applications, such as VPUs for image recognition, further increases developers' configuration, development, and management overhead.
Current cloud solutions do not yet provide simplified serverless means to use hardware processing accelerators in cloud computing environments.

We present an initial design and implementation of \name{}, an extensible and generalized serverless computing architecture that can support workloads for arbitrary hardware accelerators. 
Based on the implementation and early evaluation, we argue that serverless computing is a necessary next step in accelerated cloud computing. However, many challenges for scheduling, accelerator configuration, and workload description remain open to offering the next generation of accelerated serverless computing. 

\section*{Acknowledgment}
We thank Dr.-Ing. Anselm Busse and Michael Gebauer for supporting the initial work. This work is partly supported by the AWS Cloud Credit for Research program.

\bibliographystyle{IEEEtran}
\bibliography{references,paper}

\end{document}

%% file: src/packages.tex
\usepackage{afterpage}
\usepackage{algorithmic}
\usepackage{array} %
\usepackage{booktabs} %
\usepackage{color}
\usepackage{graphbox}
\usepackage{graphicx}
\usepackage{subcaption}
\usepackage{hyperref}
\usepackage{mathtools}
\usepackage{multirow}
\usepackage{tabularx}
\usepackage{textcomp}
\usepackage{tikz}
\usepackage{pifont} %
\usepackage{colortbl}
\usepackage{flushend}
\usepackage[normalem]{ulem}
\useunder{\uline}{\ul}{}

\usepackage[capitalise,noabbrev]{cleveref}

\usepackage{todonotes}
\usepackage{pagecolor}
\usepackage{atbegshi}

%% file: src/marcos.tex
\newcommand{\name}{\textsc{Hardless}}

%% file: src/colors.tex
\definecolor{darkblue}{rgb}{0.122, 0.435, 0.698}
\definecolor{grey}{gray}{0.30}
\definecolor{lightblue}{rgb}{0, 0.8, 0.99}

\definecolor{sky1a}{HTML}{cfe0e8}
\definecolor{sky1b}{HTML}{b7d7e8}
\definecolor{sky1c}{HTML}{87bdd8}
\definecolor{sky1d}{HTML}{daebe8}

\definecolor{sky2a}{HTML}{bccad6}
\definecolor{sky2b}{HTML}{8d9db6}
\definecolor{sky2c}{HTML}{667292}
\definecolor{sky2d}{HTML}{f1e3dd}

\definecolor{living_coral}{HTML}{FF6F61}
\definecolor{spearmint}{HTML}{64BFA4}

\definecolor{teal}{HTML}{00A08F}
\definecolor{marine}{HTML}{124653}
\definecolor{yellow}{HTML}{FEE074}
\definecolor{orange}{HTML}{FF9469}
\definecolor{pink}{HTML}{FE8d8F}

%% file: sections/00_abstract.tex
The increasing use of hardware processing accelerators tailored for specific applications, such as the Vision Processing Unit (VPU) for image recognition, further increases developers' configuration, development, and management overhead.
Developers have successfully used fully automated elastic cloud services such as serverless computing to counter these additional efforts and shorten development cycles for applications running on CPUs.
Unfortunately, current cloud solutions do not yet provide these simplifications for applications that require hardware acceleration.
However, as the development of specialized hardware acceleration continues to provide performance and cost improvements, it will become increasingly important to enable ease of use in the cloud.

In this paper, we present an initial design and implementation of \name{}, an extensible and generalized serverless computing architecture that can support workloads for arbitrary hardware accelerators. 
We show how \name{} can scale across different commodity hardware accelerators and support a variety of workloads using the same execution and programming model common in serverless computing today.

%% file: sections/01_introduction.tex
In recent years, two major trends in IT systems could be observed: an increasing use of highly specialized hardware for applications such as the Tensor Processing Unit (TPU) for AI applications and the use of highly managed and elastic cloud services such as serverless computing. 
The first trend stems from the fact that major performance improvements are no longer possible through improved general-purpose CPUs due to the breakdown of Dennard scaling in the mid-2000s~\cite{2007-Bohr-Dennard-Scalling}.
Since then, substantial performance improvements have been mainly enabled through increased specialization.
This led to the advent of hardware processing accelerators that are tailored to specific problems, including encryption, video processing, or matrix multiplication. 
The second major trend of serverless computing stems from the continued drive toward cloud automation, offloading operational tasks~\cite{2019-Kuhlenkamp-Ucc-Opstasks} to cloud vendors.
Giving customers the ability to shorten development cycles and thus enabling shorter times to market since developers can focus on the application itself rather than the underlying infrastructure~\cite{2019-Castro-ACM-Rise_of_Serverless}.
While this trend started for general computation using Function-as-a-Service offerings, the serverless model is starting to extend to other cloud computing needs.

Until now, both trends are not holistically considered in the literature.
So far, specialized solutions have emerged to meet the increasing demand for hardware-accelerated serverless applications, for example, for AI-driven systems~\cite{2019-Carrerira-Cirrus,2020-Ali-SC-Batch} but without general applicability to all types of accelerators or accelerated applications. 
Thus, we ask the research question: \textbf{How can we reduce operational tasks for accelerated applications using a serverless computing model?}

In this paper, we present \name{} as a proof of concept implementation for an extendable system that supports arbitrary accelerator solutions while promising a similar programming and execution model as other serverless systems.
Before introducing \name{} we present related work (\cref{sec:rw}), clarify the problem statement in \cref{sec:prob}. We then present the design and implementation in (\ref{sec:dgn}) of \name{} and a first experiment driven evaluation of in \cref{sec:evl} before concluding in \cref{sec:clu}.

%% file: sections/02_related_work.tex
Enabling cloud-based hardware acceleration is one of the remaining challenges in cloud computing. 
Firstly, while we can pass through a hardware device to virtualized runtimes, using the same device by multiple such runtimes is limited for hardware accelerators such as GPUs. 
High-end accelerators can be split up into different virtual devices that different virtual machines can use.
However, not all available hardware accelerators were originally designed to be used by multiple tenants. Thus, we need to resort to approaches such as multiplexing~\cite{Trimberger_1997} or partial reconfiguration~\cite{McDonald_2008}. 

A different approach to virtualizing hardware accelerator access on a single host is by distributing work across multiple smaller accelerators. 
Serverless computing is one of the emerging application patterns for creating these types of distributed applications.
Industry and academia thus are proposing approaches for serverless hardware acceleration. For instance, Zhang et al.~\cite{2021-Zhang-Faster} present a hybrid approach using serverless and serverfull systems to enable accelerated semi-serverless workloads.
Especially for machine learning, serverless acceleration has emerged as a driving factor in research~\cite{2019-Horovitz-GECON-FaaStest,2020-Ali-SC-Batch,2020-Joyner-arxiv-Ripple, 2019-Carrerira-Cirrus}. 
Operating elastic infrastructure to serve and train machine learning workloads is exceptionally costly and not recommended for sporadically used applications or models with long data collection periods.
Thus, a serverless-based machine learning application with a scale-to-zero option could be highly beneficial  \cite{2018-Carreira-Serverless-ML}. 
However, many serverless machine learning frameworks \cite{2019-Carrerira-Cirrus,2020-Ali-SC-Batch,2018-Carreira-Serverless-ML,2019-Horovitz-GECON-FaaStest,2020-Joyner-arxiv-Ripple} use the serverless system as is and would highly benefit from holistically integrated hardware acceleration to allow for larger and more complex workloads.

Accordingly, the first approaches for GPU-based serverless acceleration have appeared.
Notably, Naranjo et al.~\cite{2020-Naranjo-Accelerated} use OpenFaaS in combination with an Nvidia GPU cluster to deliver accelerated workloads. 
A similar approach by Risco et al.~\cite{2020-Risco-GPU-Enabled} propose a system named SCAR utilizing both AWS Lambda and AWS Batch, two services offered by Amazon Web Services. This approach also offloads GPU-bound tasks to a secondary service over the network.
Further, Satzke et al.~\cite{2020-Satzke-Efficient-GPU} present an architecture based on KNative named KNIX using Nvidia Docker.
We note that all these frameworks would address the needs of current machine learning applications by exposing GPUs. 
However, it is unclear if these approaches can be extended to use different kinds of hardware accelerators except GPUs.

%% file: sections/03_problem.tex
We argue that due to the growing need for accessible acceleration in cloud computing, a serverless compute architecture for processing accelerators is needed beyond the specialized solutions present so far.
Accordingly, our research question is: \textbf{How can we reduce operational tasks for accelerated applications using a serverless computing model?}

We can already infer opportunities and threats~\cite{2020-Kuhlenkamp-IC2E-ifs_and_buts} for such an accelerated serverless computing system. 
Research and industry have built workarounds to enable accelerated serverless computing~\cite{2019-Carrerira-Cirrus,2020-Ali-SC-Batch,2018-Carreira-Serverless-ML,2019-Horovitz-GECON-FaaStest} with current FaaS-Systems. However, rethinking the design of accelerated serverless applications~\cite{2021-Werner-ICSOC-ApPlCoDi} by creating a new serverless system can benefit both users and platform vendors.
Exposing heterogeneous hardware accelerators through a familiar interface and handing over the execution of work without direct input from resource consumers means that platform vendors gain the ability to fully leverage available hardware accelerators.
In addition, platform vendors can further manage where and how work is distributed to idle accelerators if we relax performance expectations.

Consequently, platform vendors can offer access to accelerator hardware at significantly less cost to service consumers, as currently under-utilized hardware can now be shared more efficiently with customers.
Similarly, offering more transparent execution environments for accelerated workloads implies that libraries and frameworks utilizing acceleration can remove existing configuration and abstraction layers, thus reducing entry barriers to developers to use acceleration. 
Lastly, by letting platform vendors mix accelerators based on user needs, specialized accelerators\footnote{e.g., the inference accelerator available at AWS: https://aws.amazon.com/machine-learning/inferentia/} can be used instead of more expensive generic accelerators as GPUs if service users want to. 
However, to enable this adaptability of accelerated applications, we must ensure that the overhead for developers remains low and that service users can clarify preferences for the type of allowed adaptation.

%% file: sections/04_design_grammerly.tex
\begin{figure}
    \centering
    \includegraphics[width=\columnwidth]{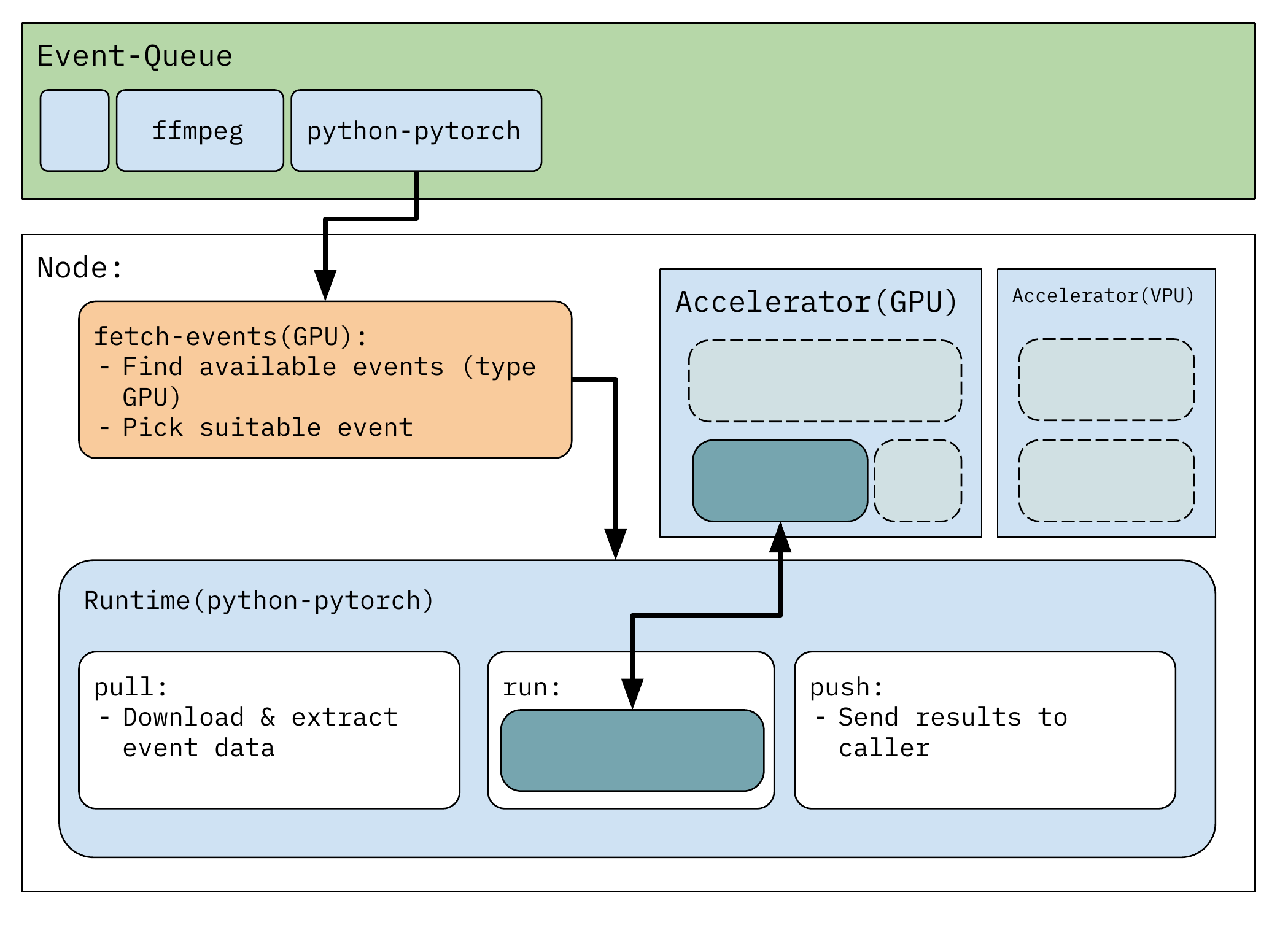}
    \caption{Abstract event flow in \name{}, showing how each node can select and run events based on locally available accelerators.}
    \label{fig:wf}
\end{figure}

\label{sec:design}
In this section, we discuss the major challenges of designing and implementing \name{}.
Specifically, we discuss the design of the runtime environments, the definition of workloads, and the implementation of the execution flow to enable accelerated serverless computing. Figure~\ref{fig:wf} shows an overview of the execution workflow used in \name{}.

\subsection{Runtime Environment}
\label{sec:design:rte}

Hardware accelerators are often accompanied by a highly specialized software stack, e.g., specific drivers, APIs, and specific programming models.
In addition, the software stacks are only qualified by the hardware vendor for specific setups both regarding hardware itself and necessary third-party software like the operating system and versions of standard libraries. 
This is contrary to the idea of serverless computing, where the developer is not supposed to be concerned with the details of the computing platform, e.g., configuring the operating system, dependencies, and file system access.
Consequently, an accelerated serverless architecture has to find a means to overcome this issue and present a simple interface to the developer no matter how the accelerators are operated.
A generalized interface also allows providers to serve user requests on different types of hardware accelerators transparently to users.
For general serverless computing, e.g., using only CPU and memory resources, developers are presented with ready-made runtime environments for specific programming languages. 
In \name{} we follow the same idea but extend this concept to accelerated software libraries, e.g., PyTorch.
In recent years, framework and library vendors have started to hide away much of the necessary programming abstraction to use accelerators from developers as long as the system is correctly configured, i.e., for a specific GPU generation, the correct drivers, and system libraries are present, e.g., CUDA v11.
For \name{} we push the responsibility to configure these systems to the platform and offer the developer a pre-configured accelerated runtime, e.g., python3-PyTorch.
Thus, reducing operational tasks from a developer's perspective.

While not all types of accelerated workloads can use such libraries and frameworks already, we argue that this trend will continue. Thus, a generalized hardware-accelerated serverless platform can use these frameworks as the main entry point of each available runtime. 
Note that with this abstraction, the developer does not need to know what kind of accelerator is actually used, and the vendor can use otherwise unused accelerators to fulfill requests.

As a foundation for our workload definition, we assume that the purpose of processing accelerators is some computational task, e.g., a function using python-PyTorch to classify images, that is performed on a data set, e.g., a set of images.
Thus, the input for a general accelerated serverless system is split into two parts: the computational task, i.e., the function definition that is unique to a workload, and the data set, which may be reused between workloads.
Like current serverless systems, we assume that workloads are stateless; that is, the computational task must fetch data sets before running, and results must be persisted elsewhere before terminating execution.
We trigger workloads using events consisting of a runtime and data set reference.
\name{} is responsible to decide how the computational task is executed on available accelerators.

\subsection{Execution Model}
\label{sec:design:execution}
We align the execution model of \name{} to the current serverless model, where events are used to invoke selected runtimes.
However, unlike conventional serverless computing, we only offer asynchronous executions, as execution times for hardware accelerators vary too much.
In the case of \name{}, an event always consists of a data set reference that needs to be fetched and additional configuration for the run method. 
A user submitting an event (data + workload reference) gets no guarantees on where and how the workload is executed.
This way, the platform is free to choose how to deploy, execute and manage all workloads.
Furthermore, this enables \name{} to scale workloads based on incoming invocations and offer similar elasticity as other computation-oriented serverless systems.

\subsection{Workflow}
\label{sec:design:workflow}
In order to support a wide variety of hardware accelerators and a variety of deployments, we also limit the execution workflow of \name{}.
Consequently, we only allow asynchronous events, as the runtime per event is unpredictable for different hardware accelerators.
A node will look for any event supported by its available accelerators, if a runtime is supported by multiple available accelerators, then the node is free to choose which accelerator to use, see Figure~\ref{fig:wf}.
Specifically, we use a distributed message queue to publish incoming events. 
Each workload can signal completion back to the event generator if the user define function allows for that. Workers do not interact with the event queue again, which enables dynamic addition and removal of worker nodes.
Further, a node can use any available accelerator since nodes are free to select workloads that they can accelerate. It is only a matter of implementing a suitable wrapper for a runtime to integrate new accelerators or add new commodity hardware into the system.

%% file: sections/05_implementation_grm.tex
\begin{figure}
    \includegraphics[width=\columnwidth]{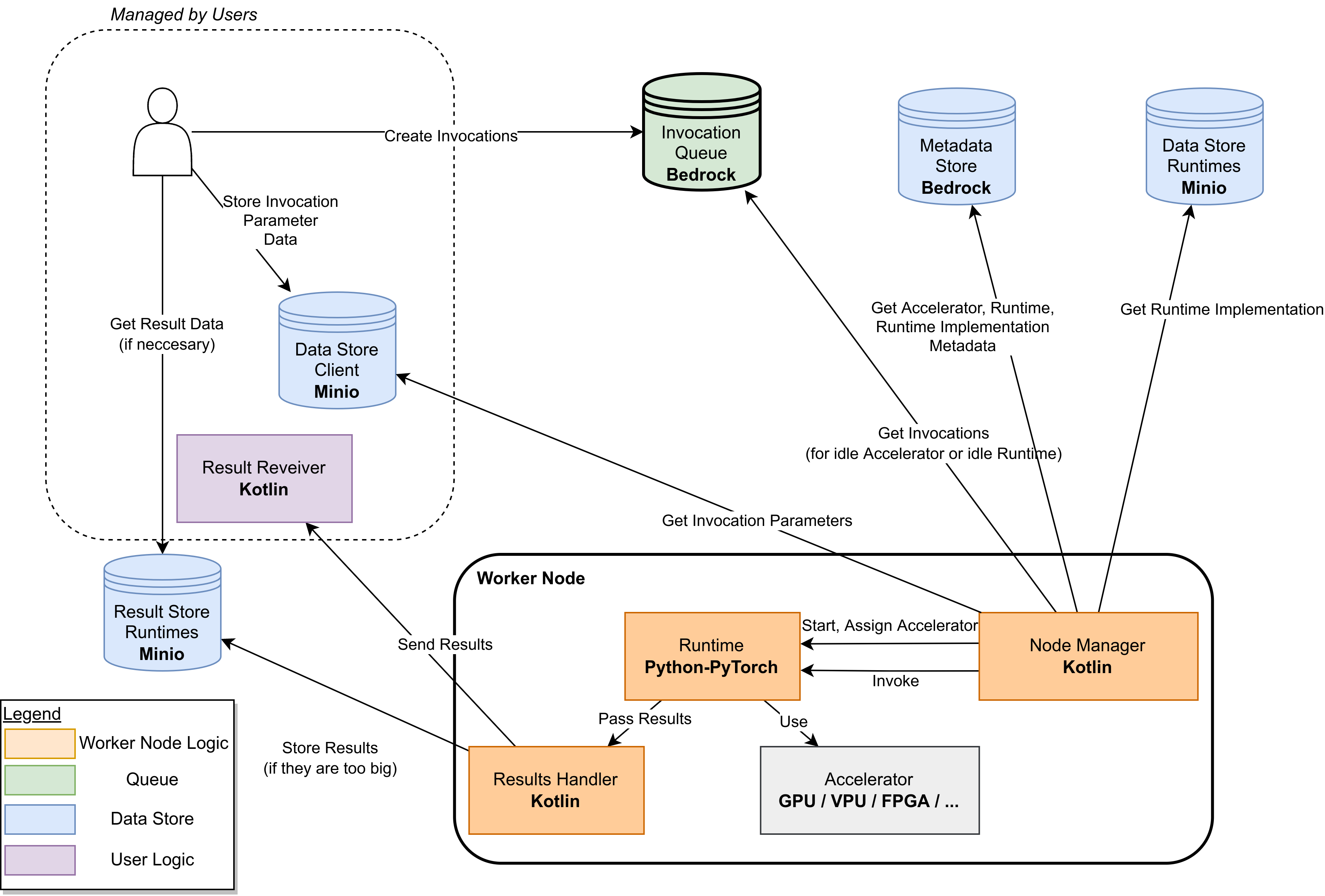}
    \caption{Detailed architecture overview of the \name{}-prototype}
    \label{fig:implementation}
\end{figure}

We implement the \name{} in Kotlin, using Bedrock for the invocation queue and Minio to store data such as runtimes.
We show that our prototype can use runtimes for ONNX Models and PyTorch. An overview of the technologies can be found in \autoref{fig:implementation}.

The node manager is responsible for managing all aspects of a single worker node, which is a machine that has access to hardware accelerators.
It starts, stops, and distributes invocations to runtime instances and assigns accelerators to them.
To perform these operations, the node manager interfaces with the invocation queue to get invocations and object storage to fetch data.
Every node manager has a list of all accelerators available to it in which it stores the type of the accelerator, a locally unique ID for it, and information necessary to schedule and balance the available resources.

A node fetches all its work from a single shared message queue. In order to allow node managers to minimize setup times and switching costs, in the serverless context typically referred to as cold-starts, the queue needs to enable each manager to scan the queue before taking invocations.
This allows worker nodes to prioritize taking workloads that are already warm.
Thus, worker nodes then need to perform two operations on this queue.
When they want to start a new workload, they need to fetch any invocation from a set of workloads that can be run on this worker node.
When an already running invocation is finished, they need to query whether the queue has invocations that have the same configuration so that the worker node can reuse an existing runtime instance.

While multiple distributed queues fulfill this requirement, we selected Bedrock\footnote{\url{bedrockdb.com/jobs}} for this prototype.

Object storage is used in this architecture to store runtime implementations, input configuration, and input data.
In our prototype, a runtime instance is a process running on a worker node that can fulfill user invocations using its runtime. We choose processes instead of containers or other forms of isolation to ensure that our system can use every type of accelerator since not all accelerators can be used with common isolation technologies such as docker.
Since the runtimes only run code that is written by the provider, security risks can be mitigated.
Different runtime instances of a runtime can be implemented to use different types of hardware accelerators, which enables our prototype to run invocations on different accelerator types transparently, however, each node needs to be configured correctly to support all available runtimes for this accelerator.

%% file: sections/06_evaluation.tex
In this section, we evaluate \name{} to show its overall functionality and how it can use arbitrary accelerators for the same user-defined workload.

\subsection{Experiment Protocol}
The goal of this evaluation is to show how \name{} can utilize different accelerators while minimizing the operational task from the perspective of a user. 
Thus, to answer this question, we compare a scaling workload while changing the availability of the accelerators in our test system.
For the experiments, we utilized an Intel Xeon E5-2630 CPU, 32GB of RAM Server, with two NVIDIA Quadro K600 GPUs, and an Intel Movidius Neural Compute Stick.

The workload model uses the same vocabulary as Kuhlenkamp et al. ~\cite{2020-Kuhlenkamp-SAC-Elasticity}. For each workload, we performed a set of invocations split into three phases (P0-P2):  a 2-minute warm-up phase (P0), a 10-minute scaling phase (P1), and a 2-minute cooldown phase (P2). Each phase has a target invocation throughput, e.g., a workload of $P0=10, P1=20, P2=20$ represents an invocation scale-up to 20 invocations per second (trps) starting with a warm-up of 10 trps.  
As a runtime, we used the established YOLO image detection model~\cite{2017-Redmon-VCPR-Yolo}, specifically the tinyyolov2.7 for ONNX. The test environment can run two parallel instances per GPU (4 in total) plus one on the Compute Stick.

As Measurmentes we collect  every invocation, we tracked its creation (RStart), when the invocation is received by a node manager (NStart),  when the execution inside the runtime starts (EStart) and stops (EEnd), when the result is received by the node manager (NEnd), and when the result is received by the benchmark client (REnd). 

We also periodically collect performance metrics in all components, including how many invocations are left in the queue (\#\textsubscript{queued}) and which accelerator is processing which event.
To aid evaluation, we calculate additional metrics based on these measurements.
In particular, we calculated the total latency of an invocation (RLat = REnd - RStart) as well as the execution time (ELat = EEnd - EStart) and delivery delay (DLat = EStart - RStart).
We also count the number of invocations that complete successfully (RSuccess).
Based on RSuccess, we then calculate a moving average number of successful computations in the last 10 seconds (\textbf{\texttt{RFast}}).

\subsection{Results}
\begin{figure}[!tbp]
    \begin{subfigure}[b]{0.5\textwidth}
        \resizebox*{\textwidth}{!}{\input{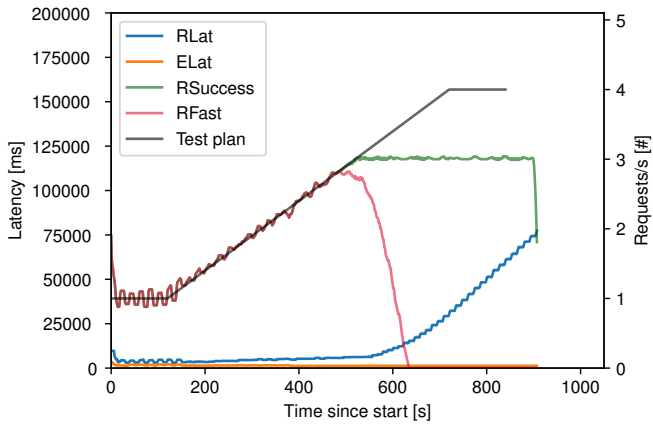}}
        \caption{Overview of complete experiment run}
        \label{eval:dualGPUWhole}
    \end{subfigure}
    \hfill
    \begin{subfigure}[b]{0.5\textwidth}
        \resizebox*{\textwidth}{!}{\input{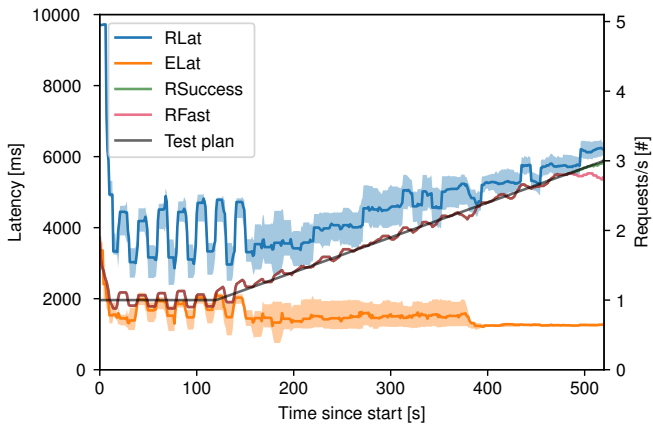}}
        \caption{Zoomed-in view, until accelerator utilization decreases}
        \label{eval:dualGPULost}
    \end{subfigure}
\caption{Client-side latency graphs for the dualGPU setup.}
\label{eval:dualGPU}
\end{figure}

\begin{figure}[!tbp]
    \begin{subfigure}[b]{0.5\textwidth}
        \resizebox*{\textwidth}{!}{\input{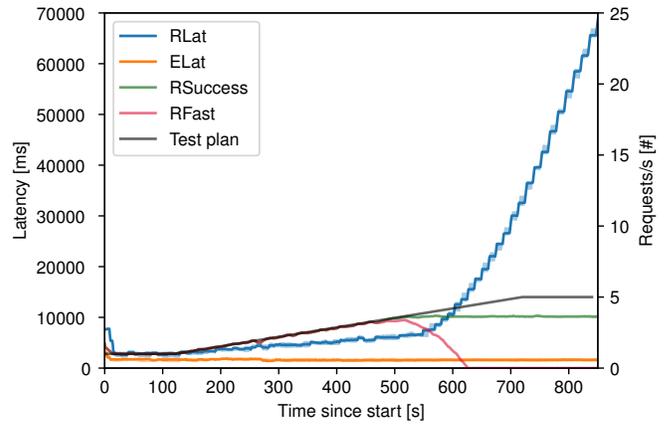}}
        \caption{Overview of complete experiment run}
        \label{eval:allVPUWhole}
    \end{subfigure}
        \hfill
    \begin{subfigure}[b]{0.5\textwidth}
        \resizebox*{\textwidth}{!}{\input{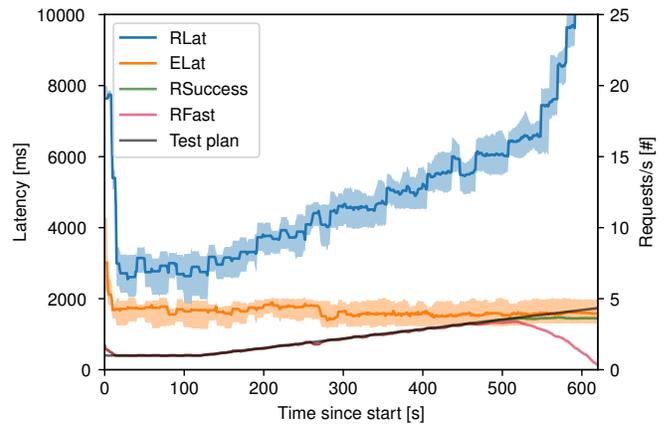}}
        \caption{Zoomed-in view, until accelerator utilization decreases}
        \label{eval:allVPULost}
    \end{subfigure}
\caption{Client-side latency graphs using all accelerators in the test environment.}
\label{eval:allVPU}
\end{figure}

We briefly present each experiment and the observed experimental results in the following. 
We observe that \name{} was able to use all accelerators available to it, resulting in a higher maximum RLat than it did with just using GPUs.
Thus, \name{} was able to utilize additional accelerators (the Intel compute stick (VPU)) without user intervention.
The experiment running just the GPUs can be seen in \autoref{eval:dualGPU}, and the experiment that uses the GPUs, as well as the VPU, can be seen in \autoref{eval:allVPU}.
As shown in \autoref{eval:dualGPULost}, the maximum RFast using two GPUs is around 3, while it is around 4 using all accelerators (cf. \autoref{eval:allVPULost}).
For the Neural Compute Stick, we observe a median ELat of 1577ms, while the median ELat for the workload running on the GPU is 1675ms.
Thus, adding the Neural Compute Stick increased the maximum RFast by about 0.75 without intervention by the service user, thus, demonstrating the platform provider's ability to utilize arbitrary idle accelerators in \name{}.

Based on these experiments, we can see that \name{} is capable of using a diverse set of accelerators without user interaction.
Further, developers can use the event-based programming model to use accelerators and benefit from a similar reduction in operational tasks as seen in serverless computing.
From a user perspective, an event is processed as fast as possible without selecting an available accelerator or configuring any hardware/software libraries.
However, the nature of hardware accelerators and the highly specialized compute stacks running on them can cause different execution times based on the provided user code. 
Customers might want specific latency or price guarantees for their invocations in a commercial setting. Thus, the platform provides and systems such as \name{} must include complex event scheduling and filtering mechanisms to ensure acceptable performance in a production setting.
Nevertheless, using the library level as a runtime for processing events allows \name{} to fully utilize all available hardware and thus ensuring less idling time for expensive resources.
However, the use of a diverse set of accelerators can quickly increase the management of platform providers. 
For example, we needed to use a much older ONNX version for the K600 GPUs that supported fewer features. Thus, events with user code might also include specific ONNX versions to enable platforms to assign work to fitting accelerators without causing reliability issues.

We argue that hardware-accelerated serverless computing is a necessary next step in cloud computing, but many challenges remain in offering cost-effective accelerated computation.